# Sub-Meter Remote Sensing of Soil Moisture Using Portable L-band Microwave Radiometer

Runze Zhang, Abhi Nayak, Derek Houtz, Adam Watts, Elahe Soltanaghai and Mohamad Alipour

*Abstract*—**Spaceborne microwave passive soil moisture products are known for their accuracy but are often limited by coarse spatial resolutions. This limits their ability to capture finer soil moisture gradients and hinders their applications in catchment hydrologic modeling and wildfire ecological impact assessment. The Portable L-band radiometer (PoLRa), compatible with tower-, vehicle- and drone-based platforms, offers soil moisture measurements from submeter to tens of meters depending on the altitude of measurement. Given that the assessments of soil moisture data derived from this portable sensor are notably lacking, this study aims to evaluate the performance of sub-meter scale soil moisture (approximately 0.7 m × 1.0 m) retrieved from PoLRa mounted on poles at four different locations in central Illinois, USA. The evaluation focuses on 1) the consistency of PoLRa measured brightness temperatures from different directions relative to the same area, and 2) the accuracy of PoLRa-derived soil moisture. As PoLRa shares many aspects of the L-band radiometer onboard the NASA's Soil Moisture Active Passive (SMAP) mission, two SMAP operational algorithms and the conventional dual-channel algorithm were applied to calculate volumetric soil moisture from the measured brightness temperatures. The vertically polarized brightness temperatures from the PoLRa are typically more stable than their horizontally polarized counterparts across all four directions. In each test period of approximately seven minutes, the standard deviations of observed dual-polarization brightness temperatures are generally less than 5 K. By comparing PoLRa-based soil moisture retrievals against the simultaneous moisture values obtained by handheld time domain reflectometry, the unbiased root mean square error (ubRMSE) and the Pearson correlation coefficient (R) are mostly below 0.04 m³/m³ and above 0.75, confirming the high accuracy of PoLRa-derived soil moisture retrievals and the feasibility of utilizing SMAP algorithms for PoLRa data. These findings highlight the significant potential of ground- or drone-based PoLRa measurements as a standalone reference for future spaceborne L-band synthetic aperture radars and radiometers. The accuracy of PoLRa yielded high-resolution soil moisture can be further improved via standardized operational procedures and appropriate tau-omega parameters.**

## I. INTRODUCTION

Accurate quantification and temporal monitoring of varied soil water content of the land surface are important across multiple disciplines, including agricultural production, natural disaster predictions, plant response to climate warming, land-atmosphere interactions, and weather forecasts [1-8]. Microwave remote sensing has emerged as an efficient and cost-effective means for estimating spatial soil moisture on a large scale [9-11]. Compared to optical and infrared sensors, microwave sensors are preferred for soil moisture mapping partly due to their subsurface sensitivity and their resistance to cloud interference [12]. More importantly, the reflective and emissive properties of surfaces at microwave frequencies correlate closely with the dielectric constant of the targeted medium, primarily influenced by the water content within it [9, 13].

In recent decades, both active and passive microwave sensors on various platforms have been extensively applied to monitor soil moisture variations at different scales, particularly by spaceborne L-band microwave radiometers [14-19]. Numerous validation studies have demonstrated the great performance of surface soil moisture products from the Soil Moisture Ocean Salinity (SMOS) and Soil Moisture Active Passive (SMAP) missions under various surface and climatic conditions [20-23]. However, due to the limitations of antenna size in spaceborne microwave radiometers, these global-scale soil moisture datasets often suffer from limited spatial resolution, typically in the tens of kilometers. This limitation hinders the capture of finer soil moisture gradients over areas of tens of square kilometers, which is often required in catchment hydrological modeling [24, 25]. Hence, various methods have been proposed to downscale coarse satellite-based soil moisture products to finer resolutions, such as 3 km, 1 km, 400 m, and even 30 m, primarily by integrating high-resolution observations from other resources [24-30]. However, these downscaled datasets often sacrifice data availability, keep inheriting the features of soil moisture in their parent pixels, and possibly introduce

This paragraph of the first footnote will contain the date on which you submitted your paper for review, which is populated by IEEE. It is IEEE style to display support information, including sponsor and financial support acknowledgment, here and not in an acknowledgement section at the end of the article. For example, "This work was supported by the United States Department of Agricultural Forest Service under grant 110789" The name of the corresponding author appears after the financial information, e.g. *(Corresponding author: Runze Zhang).*

Runze Zhang and Mohamad Alipour are with the Department of Civil and Environmental Engineering, University of Illinois Urbana-Champaign, Urbana, IL 61801 USA (e-mail: runze@illinois.edu; alipour@illinois.edu).

Abhi Nayak and Elahe Soltanaghai are with the Department of Computer Science, University of Illinois Urbana-Champaign, Urbana, IL 61801 USA (e-mail: arnayak2@illinois.edu; elahe@illinois.edu).

Derek Houtz is with Microwave Remote Sensing Group of Swiss Federal Institute for Forest, Snow, and Landscape Research, Zürich 8903 Switzerland (e-mail: derek.houtz@wsl.ch).

Adam Watts is with Pacific Wildland Fire Sciences Laboratory, United States Forest Service, Wenatchee, WA 98801 USA (e-mail: Adam.Watts@usda.gov).

additional uncertainties from supplementary data sources [31]. The challenges of further downscaling kilometer-scale soil moisture into hyper-resolution levels such as meters or even sub-meter scales are considerable. Additionally, the typical revisit frequency of spaceborne passive microwave sensors, every two to three days, is inadequate for application requiring daily updates, such as precision agriculture and commercial golf course management [32, 33]. Given the high labor costs associated with ground-based and airborne measurements, drone-based L-band radiometers have been considered an effective alternative for frequent, high-resolution tracking of soil moisture over targeted areas [34, 35].

The Portable L-band radiometer (PoLRa) is a compact, lightweight dual-polarization L-band radiometer designed for mounting on towers, poles, wheeled vehicles, and unmanned aerial systems [36]. Depending on measurement elevation, this sensor can retrieve diverse environmental parameters, including surface soil moisture and vegetation optical depth, at spatial resolutions of meters [36]. With a 3-dB full power beamwidth of 37º, the PoLRa provides dual-polarization off-nadir antenna temperatures previously unachievable by other drone-based L-band radiometers [37, 38]. Centered at 1.41 GHz, the constant incidence angle of 40º, identical to that of the L-band radiometer aboard SMAP, not only guarantees a uniform scanning footprint under the same setup but also simplifies the retrieval process by supporting a direct application of the retrieval algorithms employed in SMAP mission [36, 39].

Although PoLRa measurements have been applied for irrigation and canopy monitoring, the accuracy of their soil moisture retrievals has not been carefully examined, limiting deeper insights into the retrieval parameters and broader application. Therefore, the primary objective of this study is to conduct a comprehensive evaluation of PoLRa-derived soil moisture over bare soil and grassland. Instead of drone- or vehicle-based observations, this study focuses on sub-meter soil moisture performance (approximately 0.7 m × 1.0 m) at four sites using the PoLRa mounted on a standing steel frame setup. To the best of our knowledge, microwave passive soil moisture assessments at such a high spatial resolution have rarely been investigated. Starting with fixed-location observations helps to identify practical operational challenges of the PoLRa, thus standardizing procedures for subsequent drone-based applications. Moreover, this study also evaluates the direct use of SMAP algorithms on PoLRa-observed brightness temperatures. Currently, the default model employed in PoLRa soil moisture retrieval is the conventional dual-channel algorithm (DCA), with all the parameters set at zero (i.e., vegetation optical depth, vegetation scattering albedo, and roughness parameter) and a constant effective soil temperature of 292.15 K, considering the lack of reliable ancillary products at such a high resolution. SMAP's operational algorithms, namely the single-channel algorithm (SCA) and the regularized dual-channel algorithm (RDCA), with their paired parameters, have also been utilized to assess the applicability of SMAP algorithms for PoLRa [39, 40]. Though straightforward, this initial validation of PoLRa-based sub-meter soil moisture retrievals establishes a solid foundation for enhancing the quality of soil moisture derived from drone-based PoLRa and demonstrates the prominent potential of using ground-/drone-based PoLRa-retrieved areal soil moisture as an alternative reference for airborne- and spaceborne observations, relative to the *in-situ* sensors that can only detect wetness at a point scale.

The paper is structured as follows. In Section 2, we describe the details of the timeline and experimental layout for collecting PoLRa observations as well as the steps of brightness temperature pre-processing and soil moisture retrieval flows with validation metrics. Then we present the results and discuss the limitations and future work in Section 3. Finally, conclusions are presented followed by a summary in Section 4.

## II. Data and Methods

### A. Study sites and experimental timeline

The experiment was structured into three phases, categorized by testing sites and experimental dates as depicted in Figure 1. Phase one (Nov 3rd – 4th, 2023) and phase two (Nov 7th – 8th, 2023) involved field tests over two bare-soil locations within a cornfield in central Illinois (Figure 1a and c), but were prematurely halted due to harvesting activities. The study area for phase three (Nov 11th – 25th, 2023) was then shifted to a bare-soil location and a grassland plot on the campus of the University of Illinois Urbana-Champaign (Figure 1d). The setup of the measuring instrument, positioned 1.14 m above the ground, maintained a constant elliptical footprint with major and minor axes of 1.4 m and 0.7 m, respectively. The largest square inscribed in this ellipse was approximated, with its center and corners designated as benchmark points (Figure 1b). Soil volumetric water content at these five points was simultaneously measured by a handheld time domain reflectometry (TDR) sensor (METER TEROS 12: https://metergroup.com/products/teros-12/teros-12-tech-specs) across three phases, and their spatial average was used as the referenced soil moisture for assessing the accuracy of PoLRa-based soil moisture retrievals. Key measuring locations, including benchmark points and the two front wheels of the sensor stand, were marked with pins to ensure consistent testing areas. Each measurement session began with a five-minute warm-up period after starting the PoLRa. During phases one and two, measurements were taken from four directions encircling the same elliptical area (Figure 1b). In phase three, however, measurements were limited to one direction partly due to pedestrian and vehicular activity. Measurement times were consistently scheduled from 2 pm to 4 pm to minimize diurnal variations affecting the results.

### B. Pre-processing of observed brightness temperatures

Given that the PoLRa's sampling frequency across its four switch positions is approximately 69 milliseconds, thousands of dual-polarization brightness temperatures can be collected during each five to seven-minute measurement interval [36]. Initial data analysis revealed occasional outliers, likely due to improper operations or external interference. Thus, a series of pre-screening steps were employed before incorporating these brightness temperatures into soil moisture retrieval algorithms (Figure 2). In accordance with the SMOS Level 1 manual, the maximum threshold was preset as 320 K for brightness temperature at both polarizations. Meanwhile, the minimum thresholds were computed using the forward tau-omega radiative transfer model by assuming volumetric soil moisture of 1 m$^3$/m$^3$ [39]. A constraint that the vertically polarized brightness temperature is larger than its coincident horizontal polarized brightness temperature was additionally applied. Although it is possible to generate a time series of soil moisture from those filtered brightness temperatures, we only preserved one representative brightness temperature per measurement session for simplicity. We presumed that soil moisture values averaged from the time series and derived from the representative brightness temperature have similar magnitudes. The use of the median as the representative brightness temperature was determined by an investigation that compared the 25$^{th}$, 50$^{th}$, and 75$^{th}$ percentiles and mean values of all the brightness temperatures from phase three.

*C. Soil moisture retrieval algorithm*

Since PoLRa shares many aspects of the SMAP radiometer, the tau-omega microwave radiative transfer model designed for SMAP soil moisture retrieval was employed here to estimate volumetric soil moisture from PoLRa observations [39, 41]. This model grouped the radiation intensity emitted from the land surface into three components: 1) upward soil emission attenuated by the overlying canopy, 2) upward vegetation emission, and 3) downward vegetation emission reflected by the soil surface and subsequently attenuated by the canopy, after excluding atmospheric, cosmic and galaxy contributions (Equation 1). The specific inversion process, illustrated in Figure 2, starts with calculating effective emissivity by normalizing observed brightness temperature using a soil effective temperature approximated by the weighted average of soil temperatures at the surface and 10 cm depth [39, 42]. Here, soil effective temperature is assumed to be numerically equal to the coincident soil temperature measured by the TDR sensor with a 5.5 cm tine. The effective soil emissivity is adjusted for vegetation contribution using the vegetation optical depth ($\tau$) and scattering coefficient ($\omega$) [39, 41]. The roughness effect is often quantified through semi-empirical models to yield the soil emissivity from the smooth soil [43, 44]. Using the Fresnel equations, the smooth soil emissivity is then converted into the complex dielectric permittivity of the bulk smooth soil-water-air system [39]. A dielectric mixing model, such as the Mironov model, can be subsequently used to transform the dielectric constant to volumetric (or gravimetric) soil moisture [45-48]. Due to the complexity of inverting all these sub-models, in practice, soil moisture retrievals usually rely on an optimization procedure that minimizes the differences between brightness temperatures observed by radiometers and simulated by the forward model.

$$T_{B_p} = \gamma e_p T_e + (1 - \omega)(1 - \gamma)T_e$$
$$+ \gamma(1 - e_p)(1 - \omega)(1 - \gamma)T_e \quad (1)$$

where the subscript p refers to the polarization, $\gamma$ is the transmissivity of the overlying vegetation layer (as a function of $\tau$ and incidence angle), $e_p$ denotes the rough soil emissivity, $T_e$ represents the soil effective temperature, and $\omega$ is the effective vegetation scattering albedo.

Depending on the treatment of vegetation radiation, the operational SMAP algorithms are categorized into the single-channel algorithm (SCA) and the regularized dual-channel algorithm (RDCA). In the SCA approach, the annual climatology of vegetation opacity is quantified through the observed normalized difference vegetation index (NDVI), and then soil moisture is estimated using brightness temperature from a single polarization [39]. On the other hand, RDCA, which is derived from the modified dual-channel algorithm, utilizes the brightness temperatures from both polarizations to simultaneously estimate soil moisture and vegetation opacity, with an additional regularized term in its cost function (Equation 4) [40].

Besides the structural aspects of the algorithms, parameters such as the roughness parameter and vegetation scattering can also influence the accuracy of soil moisture retrievals [43, 49-52]. For SMAP, roughness and vegetation scattering parameters remain temporally static and uniform within a given 1 km land cover class [39]. In the context of observations at hyper-resolutions finer than 1 km, however, these parameters are rarely available. This absence can be especially critical when attempting soil moisture retrieval at meter-scale resolutions, particularly in areas with diverse land covers. To gauge the impact of insufficient ancillary information, the DCA where the parameters and temperature arbitrarily assumed as 0 and 292.15 K is included in the analysis. Table 1 details the configuration for each algorithm employed in PoLRa-observed soil moisture retrieval, with the DCA further categorized as DCA0, DCA1, and DCA2 based on different parameters, temperature, and dielectric model scenarios. However, it is worth noting that while directly using SMAP parameters from land cover look-up tables is not deemed optimal, it at least serves as a reasonable first guess for reference.

$$F_{SCA}(sm) = \left[ T_{B_p}^{sim}(sm) - T_{B_p}^{obs} \right]^2 \quad (2)$$

$$F_{DCA}(sm, \tau) = \left[ T_{B_V}^{sim}(sm, \tau) - T_{B_V}^{obs} \right]^2$$
$$+ \left[ T_{B_H}^{sim}(sm, \tau) - T_{B_H}^{obs} \right]^2 \quad (3)$$

$$F_{RDCA}(sm, \tau) = \left[ T_{B_V}^{sim}(sm, \tau) - T_{B_V}^{obs} \right]^2$$
$$+ \left[ T_{B_H}^{sim}(sm, \tau) - T_{B_H}^{obs} \right]^2$$
$$+ \lambda^2(\tau - \tau_{SCA})^2 \quad (4)$$

where F represents the cost function for the soil moisture retrieval algorithm denoted by the subscript. sm refers to trail soil moisture used to forward simulation of brightness temperature, and $\tau$ is the vegetation opacity where $\tau_{SCA}$ is estimated using NDVI. $\lambda$ is a regularized parameter equal to 20.

To estimate grassland soil moisture during phase three using SCA, it is essential to derive vegetation opacity from NDVI. For this purpose, 30-meter surface reflectance at the red (SR_B4) and near-infrared band (SR_B5) from the Landsat 9 Level 2 dataset are used, specifically from November 2023 [53]. Observations affected by clouds were excluded using appropriate masking. NDVI images for four discrete dates - Nov 4th, 11th, 20th, and 27th - were calculated over the targeted spot. These images were then linearly interpolated to generate a daily time series of NDVI. These NDVI estimations were converted into vegetation opacity using the formulas outlined in [39]. Additionally, clay fraction, another vital input for the Mironov's model, was obtained from the SoilWeb database by focusing on the primary component within the soil over the testing site [54].

The accuracy of soil moisture retrievals is often described by four conventional metrics: bias, root-mean-square error (RMSE), unbiased root-mean-square error (ubRMSE), and Pearson correlation (R) [55]. These metrics could effectively reflect the discrepancies in terms of magnitudes as well as the temporal correlation between the referenced and testing soil moisture products. The formulas used to compute these metrics are shown in Equations (5) to (8).

$$\text{bias} = \text{E}[\text{sm}_{\text{obs}}] - \text{E}[\text{sm}_{\text{ref}}] \tag{5}$$

$$\text{RMSE} = \sqrt{\text{E}[(\text{sm}_{\text{obs}} - \text{sm}_{\text{ref}})^2]} \tag{6}$$

$$\text{ubRMSE} = \sqrt{\text{RMSE}^2 - \text{bias}^2} \tag{7}$$

$$\text{R} = \frac{\text{E}[(\text{sm}_{\text{obs}} - \text{E}[\text{sm}_{\text{obs}}])(\text{sm}_{\text{gnd}} - \text{E}[\text{sm}_{\text{ref}}])]}{\sigma_{\text{obs}}\sigma_{\text{ref}}} \tag{8}$$

where E […] represents the arithmetic mean; the subscript obs and ref denote SM retrievals from PoLRa observations and the spatially averaged TDR measurements; $\sigma$ refers to the standard deviation of the SM retrievals.

## III. RESULTS AND DISCUSSION

### A. Investigation on PoLRa-observed brightness temperature

For phases one and two, the consistency of PoLRa observations was assessed by comparing brightness temperatures from four directions over the same area. Figure 3 describes boxplots of filtered brightness temperatures at both horizontal and vertical polarizations. Vertical polarizations show relative stability across directions compared to horizontal ones, where larger deviations in horizontally polarized brightness temperatures are partly attributed to its increased sensitivity to surface roughness. Within each box, fluctuations are generally minor, supported by their small standard deviations mostly below 5 K. Preliminary analysis of DCA0 soil moisture on Nov 11th indicates deviation at this extent could lead to soil moisture fluctuations around $\pm$ 0.015 $m^3/m^3$. There are a few abnormal boxes, such as on Nov 8th (Figure 3d), likely because of improper operations. It was observed that keeping the laptop on and attaching the PoLRa to it via an ethernet cable during measurements could significantly increase observed raw voltages, which will be subsequently converted into the brightness temperature and soil moisture, with substantial noise (Figure S1).

The criteria for selecting representative brightness temperatures during phase three involved comparing 25th, 50th, and 75th percentiles and mean values of collected brightness temperatures (Figure 4). Again, brightness temperatures from one test are typically clustered tightly. Variations in vertical polarizations are less pronounced than in horizontal. Over time, brightness temperatures at bare soil and grassland locations are consistent, though non-vegetated areas showed more pronounced fluctuations. Given the minor differences between mean and median values, the median brightness temperatures, the median brightness temperature was chosen as the representative value for each test period over a given area.

### B. Performance of PoLRa-derived soil moisture

Statistical metrics that reflect the accuracy of PoLRa-derived soil moisture across different phases are listed in Table 2. The ubRMSE values are generally below 0.04 $m^3/m^3$, meeting the SMAP mission requirements [56]. Comparing soil moisture derived from three algorithms under six scenarios shows similar but complementary performances at different locations, highlighting the necessity of incorporating multiple algorithms retrieving high-resolution soil moisture, especially in the absence of ancillary information. Correlation coefficients (i.e., R) mostly exceed 0.75, indicating strong temporal consistency between PoLRa-derived and TDR-measured soil moisture. The high R values for the continuous two-week observations confirm PoLRa's adequacy to track soil moisture trends at hyper-resolution.

Notably, the performance of PoLRa-based sub-meter soil moisture retrievals seems substantially contingent on testing sites. The ubRMSE values at the bare-soil spot in the University of Illinois Urbana Champaign commonly exceed 0.045 $m^3/m^3$ whereas those in phase one location are typically smaller than 0.01 $m^3/m^3$. This discrepancy may be due to the geographic compatibility of vegetation scattering and roughness parameters used in the retrieval process or due to the relatively low magnitude Radio Frequency Interference (RFI) present around the university campus. In terms of land cover, soil moisture variations in bare areas align more closely with TDR measurements than those in grassland.

Time series of soil moisture retrievals from different algorithms and TDR measurements for phase three are shown in Figure 5 where the light green patched areas surrounding the TDR data represent the range of two standard deviations. No data was available on Nov 20th due to rainfall. The varying amplitudes of PoLRa-derived soil moisture over the bare area are notably higher than those over grassland (Figure 5). However, such differences are not observed in TDR measurements. DCA0 soil moisture occasionally exhibits the opposite trends relative to others (e.g., Figure 5a), likely due to ignoring temporal variations in soil effective temperatures. Nevertheless, this divergence cannot be simply regarded as an error, as variations in TDR soil moisture at benchmark points do not always align perfectly (Figure S2). Moreover, PoLRa-based retrievals tend to overestimate soil moisture systematically, likely due to the use of raw TDR-measured soil temperature (dashed orange lines in Figure 5). This overestimation might be corrected by modifying the measured soil temperature with a positive offset.

*C. Discussion*

While SMAP algorithms effectively derive sub-meter soil moisture from PoLRa-observed brightness temperature, the quality of these retrievals can be compromised by the lack of suitable parameters at the correspondent spatial resolution. Previous studies indicates that vegetation scattering effects and roughness parameters could vary significantly across time and within the same land cover span of tens of kilometers, and more so at meter scales [44, 57]. When mapping soil moisture from drone-based PoLRa observations over large areas, using the correct parameters is crucial for accurately representing the spatial patterns of soil moisture. High-resolution ancillary data for soil moisture retrieval from ground- or drone-based microwave radiometry can be enhanced by integrating Lidar, optical, near- and thermal-infrared sensors mounted on the unmanned aerial system alongside PoLRa detection [34]. Alternatively, soil moisture data measured by handheld probes over small sections of targeted areas can be used to calibrate vegetation scattering and roughness parameters in near real-time across land cover types. Assuming these parameters are constant in each land cover class within the targeted area, they can be applied to the remaining areas of their respective land cover.

Moreover, the PoLRa-based brightness temperatures in this study may be impacted by imperfect operations and external interference. For instance, during phases one and two, the measurements were not taken continuously between different directions and not recorded under the same data file, potentially undermining the stability achieved through the initial warm-up procedures by frequently restarting. Additionally, the presence of an antenna tower near the phase three locations might have introduced signal interference. Future work is needed to understand and characterize the effects of electromagnetic interference on the performance of the instrument. The conversion of raw voltages from the PoLRa sensor to L-band brightness temperatures involves using two offsets typically determined by preliminary cold-sky calibration [36]. During the experiment, however, these parameters were only calibrated once for phase three, and offsets with the default setting at 0 were used for phases one and two. In the process of performing ground-based PoLRa field tests, the specific experimental steps have been gradually adjusted to pursue observations with better quality. However, some discrepancies from these changes may not be completely omitted.

Although the experimental sites cover four distinct spots with two different land covers, the limited number of tests may not fully capture the performance of PoLRa-derived soil moisture retrievals under all conditions. As soil wetness affects the measurement depth of the L-band radiometer, the effective depths at which PoLRa-derived soil moisture and TDR measurements are taken do not always align [58]. Future investigations will expand drone-based PoLRa observations to broader areas with varied vegetation. Since the L-band radiometer detects the radiation intensity from an integrated soil-vegetation system, the improved accuracy of soil moisture would also facilitate the precise retrieval of vegetation optical depth. The microwave vegetation optical depth has been widely used as a proxy of vegetation water status as well as acting as a wildfire indicator for live fuel moisture [59, 60]. To improve the retrieval quality of water content within both soil and vegetation, local parameterization strategies will be explored alongside drone-based PoLRa measurements in future work. Besides capturing diverse environmental parameters, another promising use of the PoLRa involves leveraging its aggregated dual-polarization brightness temperatures as an independent reference. This reference can be used for comparison and calibration with the upcoming L-band NASA-ISRO SAR (NISAR) and ESA ROSE-L missions, which aim to deliver high-resolution soil moisture data at approximately 100 m resolution [61, 62].

## IV. Conclusion

This study evaluates sub-meter soil moisture retrievals (approximately 0.7 m × 1.0 m) from pole-fixed PoLRa at four sites in central Illinois. Conducted in November 2023, the experiments were divided into three phases based on the timeline and measurement area. Brightness temperatures recorded by PoLRa during each seven-minute test were first filtered using arbitrarily defined thresholds. The median value of these filtered brightness temperatures was selected as the representative value for the duration of the observation. Three different algorithms, including two operational SMAP algorithms, were employed under six scenarios to estimate the volumetric soil moisture from these representative brightness temperatures. The accuracy of these soil moisture estimates was then validated by comparison with the average soil moisture from five uniformly distributed TDR-measured benchmarks within the targeted pixel.

During the first two phases, the brightness temperatures measured, in four directions by the PoLRa instrument displayed similar magnitudes, with vertically polarized brightness temperatures showing more stability than the horizontally polarized counterpart. The variation in filtered brightness temperature from one test typically has a standard deviation below 5 K, demonstrating the stable measurement of radiation intensity by PoLRa. The average ubRMSE value for PoLRa-derived soil moisture retrievals is less than 0.04 m³/m³, which is in line with the accuracy requirements of the SMAP mission. With R values generally above 0.75, PoLRa's measurements reliably captured the temporal variability of soil moisture, even at a one-meter scale.

While SMAP models and the DCA algorithm can derive soil moisture from PoLRa observations, no single algorithm consistently outperforms others. Additionally, employing one kilometer-scale vegetation scattering and roughness parameters (from SMAP) or omitting these effects could introduce errors and degrade the retrieval accuracy. PoLRa-based retrievals tends to overestimate oil moisture compared to TDR measurements, perhaps due to the inappropriate use of TDR-measured soil temperature directly in the retrieval flows. It is anticipated that the accuracy of PoLRa-based soil moisture retrievals could be improved in future tests through more regularized experimental operations and the use of reliable, high-resolution ancillary data. Strategies for local parameterization in the tau-omega microwave radiative transfer model will be examined and applied to drone-based PoLRa observations over broader areas.

ACKNOWLEDGMENT

This study was funded by the United States Department of Agriculture Forest Service under grant 110789. The authors sincerely appreciate this support.

FIGURES AND TABLES

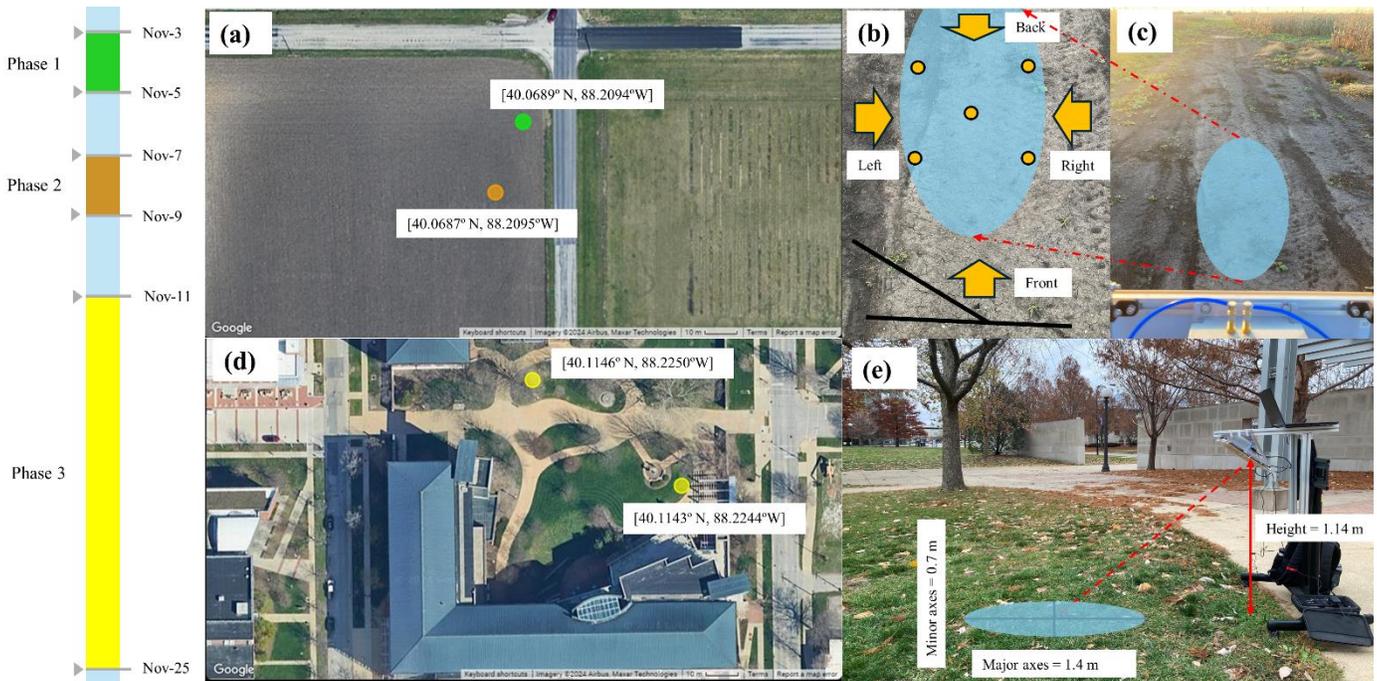

**Figure 1.** Experimental sites, timeline, and instrument setup of ground based PoLRa measurements: (a) geographical coordinates of experimental locations for phase one and two; (b) microwave radiation measurements over the same area from four different directions and five TDR measurement points; (c) the footprint area observed during phase one; (d) geographical coordinates of experimental locations for phase three; and (e) the setup and geometric view of ground based PoLRa instruments.

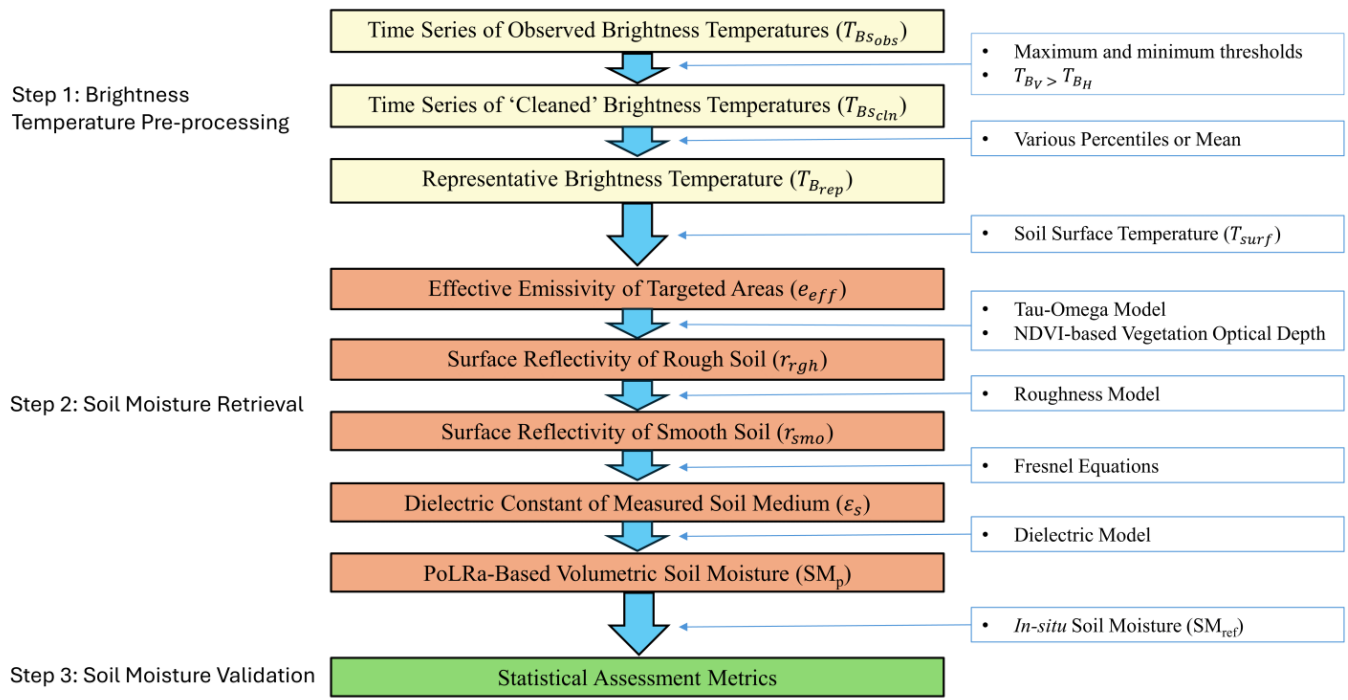

**Figure 2.** Process flow chart that describes the conversion of PoLRa-derived brightness temperatures to areal soil moisture over the targeted locations with performance assessment.

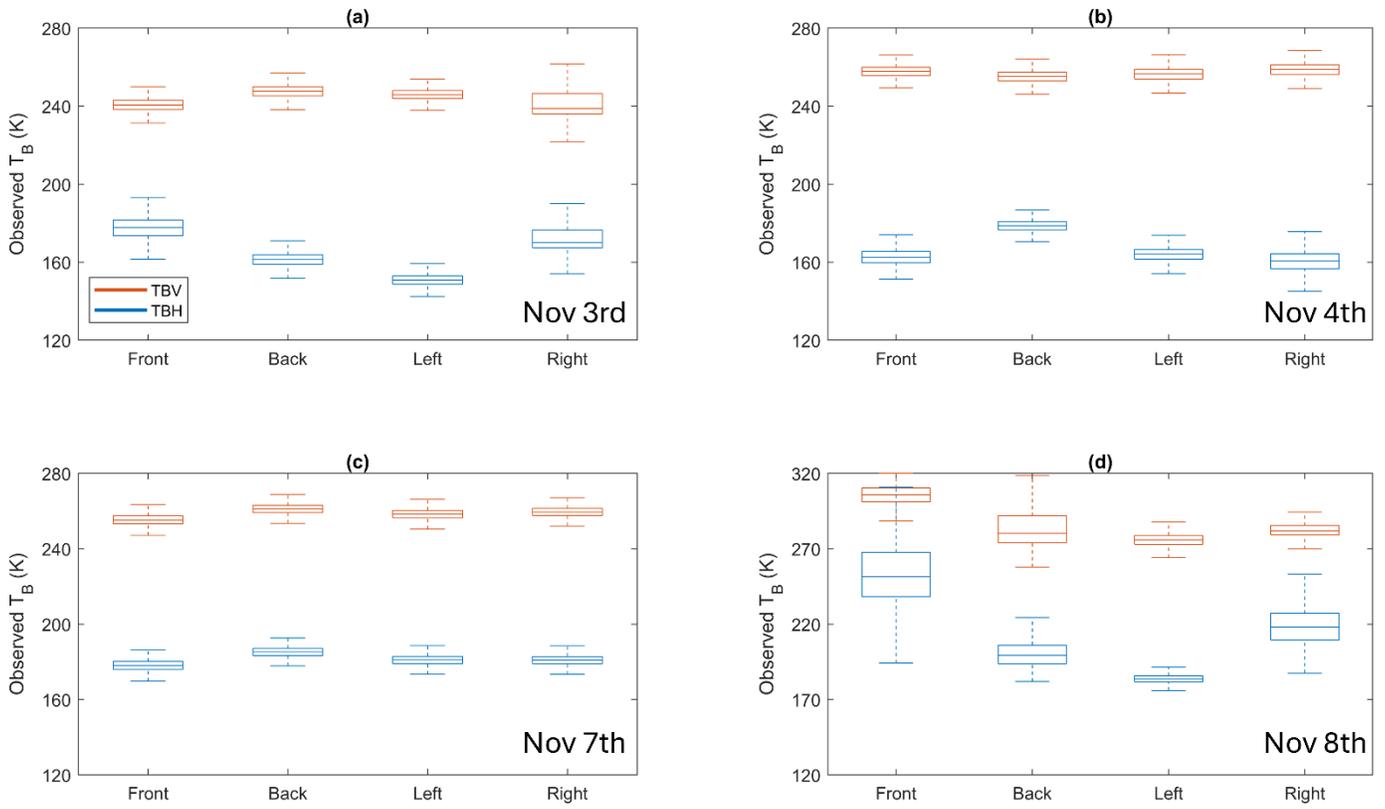

**Figure 3**. Boxplots of polarized brightness temperatures over the testing sites from four different directions on four dates: (a) November 3rd, 2023; (b) November 4th, 2023; (c) November 7th, 2023; and (d) November 8th, 2023.

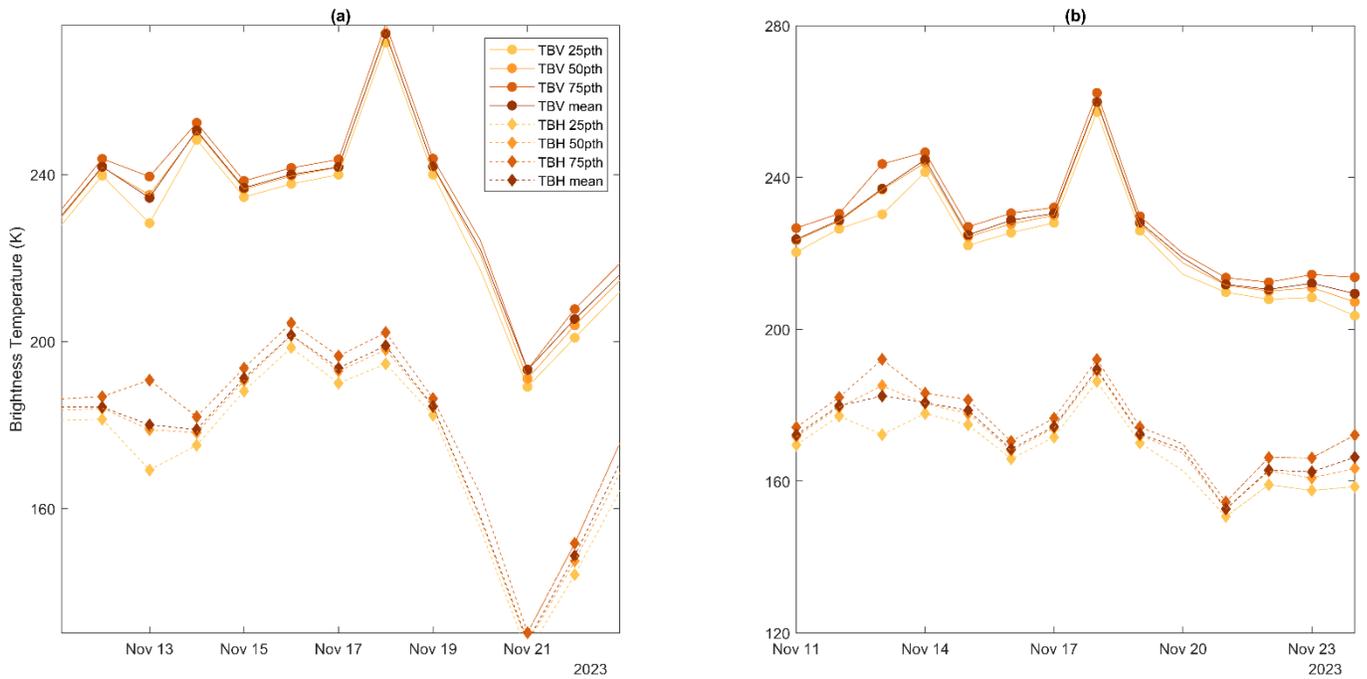

**Figure 4**. Time series of different representative polarized brightness temperatures extracted from the daily sets of filtered brightness temperatures during phase three over (a) bare soil and (b) grassland.

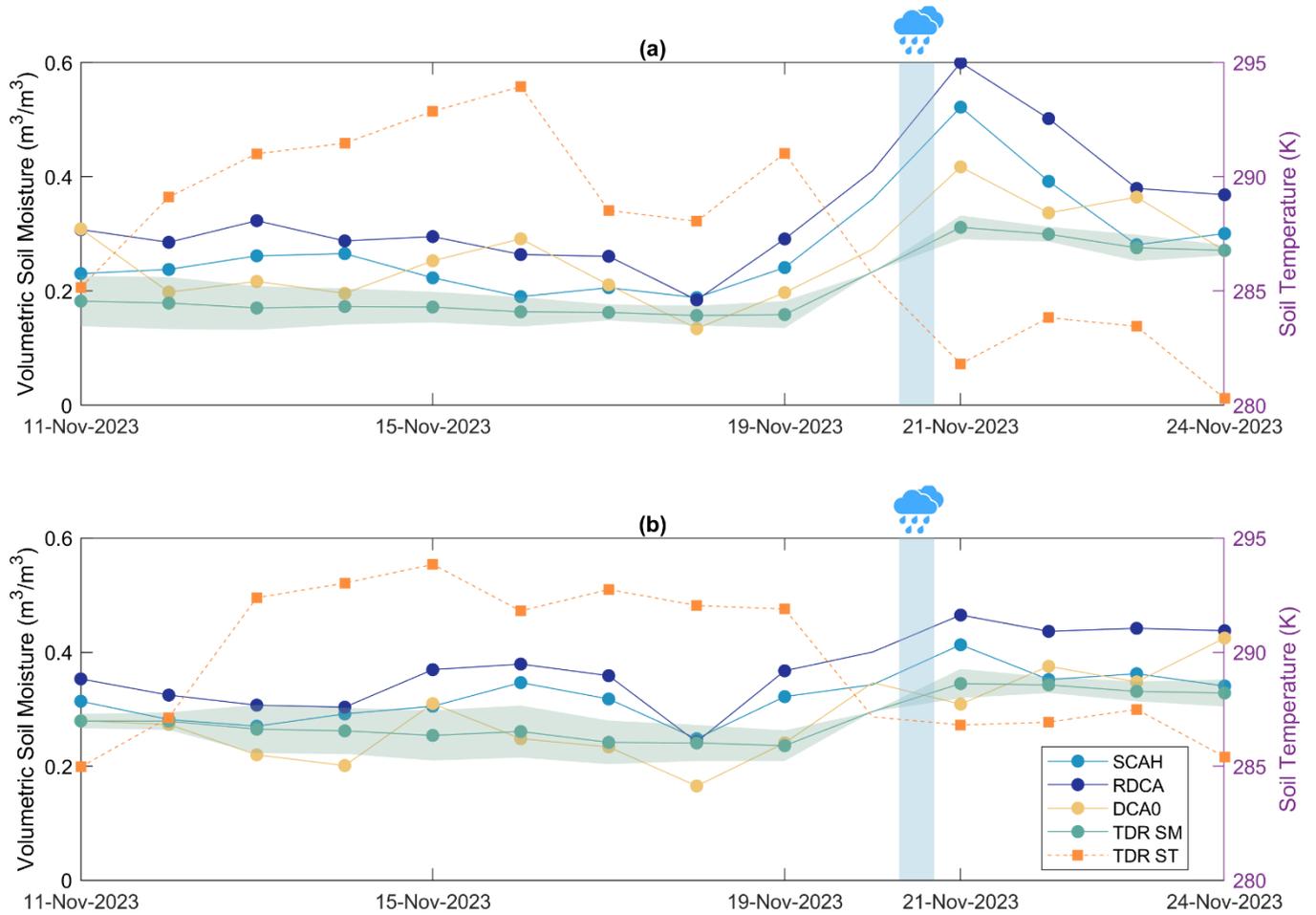

**Figure 5**. Time series of soil moisture derived from PoLRa-derived brightness temperatures using three different algorithms (SCAH, RDCA and DCA0) and measured by the TDR instrument (TDR) during phase three over (a) bare soil and (b) grassland.

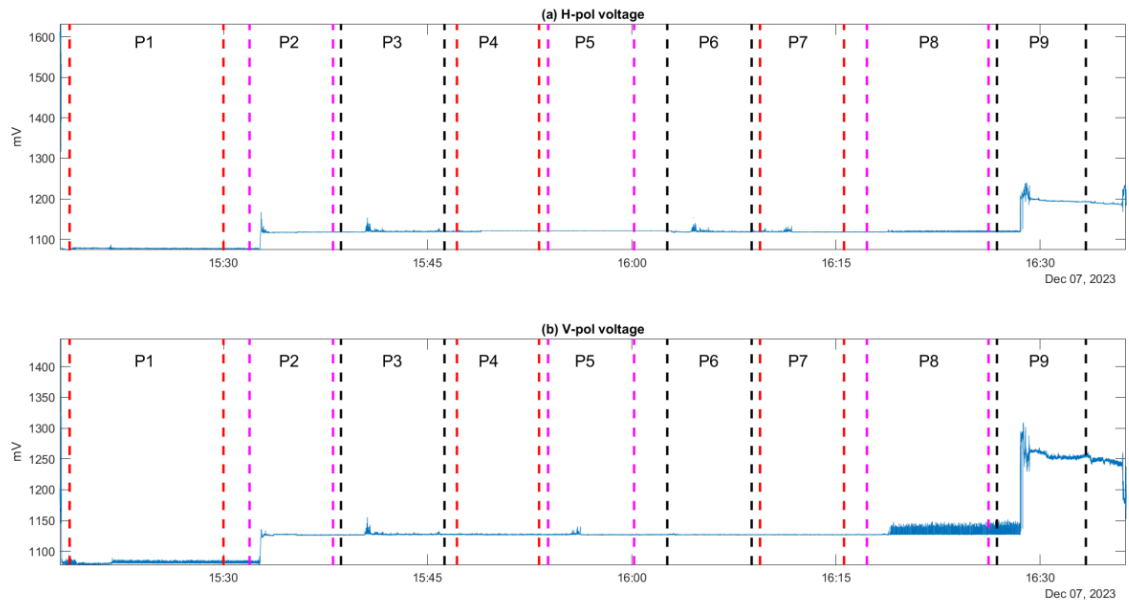

| Phase | Target Area | Ethernet Cable Connection | Laptop Status | RF Cable Connection |
|-------|-------------|---------------------------|---------------|---------------------|
| 1 | Cold Sky | Detached | Off | Both Connected |
| 2 | Ground | Detached | Off | Both Connected |
| 3 | Ground | Detached | On | Both Connected |
| 4 | Ground | Detached | On | Left Connected |
| 5 | Ground | Detached | Off | Left Connected |
| 6 | Ground | Detached | On | Right Connected |
| 7 | Ground | Detached | Off | Right Connected |
| 8 | Ground | Attached | Off | Both Connected |
| 9 | Ground | Attached | On | Both Connected |

**Figure S1**. Time series of raw voltages measured by PoLRa dual-polarization antenna with different settings where the details are attached in the table below.

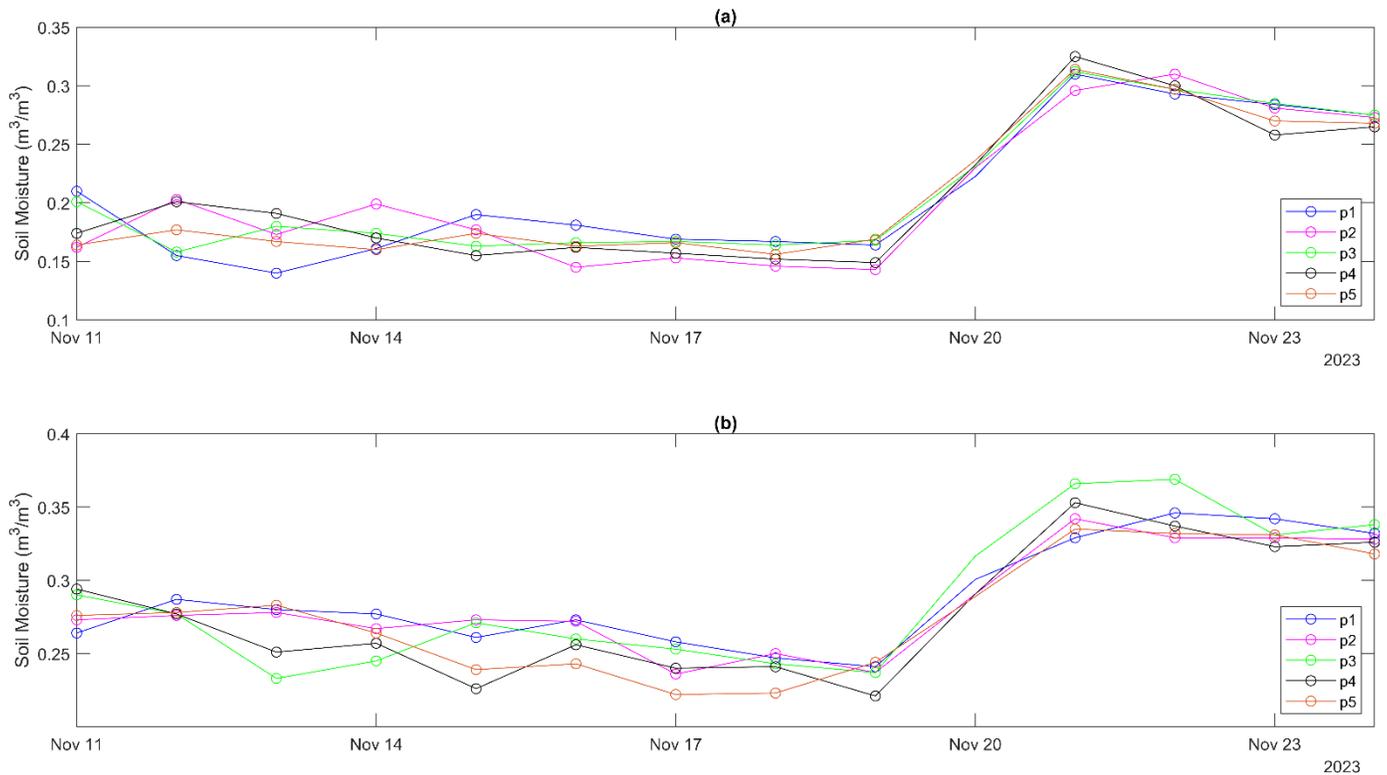

**Figure S2**. Time series of volumetric soil moisture measured by the TDR instrument from five points within the same areas of (a) bare soil and (b) grassland during phase three.

TABLE I
SOIL MOISTURE RETRIEVAL ALGORITHMS AND CONFIGURATIONS CONSIDERED IN THIS STUDY

| Algorithm\Configuration | Polarization | Roughness ($h$) | Single Scattering Albedo ($\omega$) | Soil Effective Temperature ($T_e$) (K) | Nadir Vegetation Optical Depth ($\tau_{nad}$) | Dielectric Model |
|---|---|---|---|---|---|---|
| SCAV | Vertical | 0.15 (bare soil) 0.156 (grassland) | 0 (bare soil) 0.05 (grassland) | $T_s{}^*$ | NDVI-based Estimation | Mironov 2009 |
| SCAH | Horizontal | 0.15 (bare soil) 0.156 (grassland) | 0 (bare soil) 0.05 (grassland) | $T_s{}^*$ | NDVI-based Estimation | Mironov 2009 |
| RDCA | Both | 0.4612 | 0 (bare soil) 0.0608 (grassland) | $T_s{}^*$ | Output | Mironov 2009 |
| DCA0 | Both | 0 | 0 | 292.15 | Output | Topp 1980 |
| DCA1 | Both | 0 | 0 | 292.15 | Output | Mironov 2009 |
| DCA2 | Both | 0 | 0 | $T_s{}^*$ | Output | Mironov 2009 |

$^*$ $T_s$ denotes soil temperature measured by the handheld time domain reflectometry sensor.

TABLE II
STATISTICAL ASSESSMENT METRICS OF PoLRa SOIL MOISTURE RETRIEVALS

| | ubRMSE (m³/m³) | | | | | |
|---|---|---|---|---|---|---|
| | SCAV | SCAH | RDCA | DCA0 | DCA1 | DCA2 |
| Phase 1 | 0.008 | 0.011 | 0.000 | 0.007 | 0.007 | 0.002 |
| Phase 2 | 0.041 | 0.029 | 0.040 | 0.025 | 0.024 | 0.025 |
| Phase 3 – Soil | 0.046 | 0.050 | 0.059 | 0.046 | 0.046 | 0.046 |
| Phase 3 - Grass | 0.038 | 0.030 | 0.039 | 0.045 | 0.046 | 0.041 |
| | R | | | | | |
| | SCAV | SCAH | RDCA | DCA0 | DCA1 | DCA2 |
| Phase 1 | 1.000 | -1.000 | 1.000 | 1.000 | 1.000 | 1.000 |
| Phase 2 | 1.000 | 1.000 | 1.000 | 1.000 | 1.000 | 1.000 |
| Phase 3 – Soil | 0.829 | 0.856 | 0.899 | 0.809 | 0.809 | 0.746 |
| Phase 3 - Grass | 0.743 | 0.719 | 0.794 | 0.805 | 0.806 | 0.757 |